\documentclass[aps,prl,twocolumn,showpacs,preprintnumber,amsmath,amssymb,superscriptaddress]{revtex4}
\usepackage{graphicx}
\usepackage{dcolumn}
\usepackage{bm}

\def\LNpi{$\Lambda \to N\pi$}
\def\LN{$\Lambda N \to NN$}

\def\Lnn{$\Lambda n\to nn$}
\def\Lnp{$\Lambda p\to np$}
\def\pik{($\pi^+$,$K^+$)}

\def\Gn{$\Gamma_n$}
\def\Gp{$\Gamma_p$}
\def\Gnp{$\Gamma_n/\Gamma_p$}
\def\Gnn{$\Gamma_{2N}$}
\def\pim{$\pi^-$}

\def\HefiveL{$^{5}_{\Lambda}$He}
\def\He5lam{$^{5}_{\Lambda}$He}
\def\Li6lam{$^{6}_{\Lambda}$Li}
\def\Clam{$^{12}_{\Lambda}$C}
\def\Lisix{$^{6}$Li}

\def\He5lam{$^5_\Lambda$He}

\def\e{\epsilon}

\begin{document}
\preprint{APS/123-QED}
\title{Exclusive Measurement of the Nonmesonic Weak Decay of \HefiveL~Hypernucleus}

\author{B.~H.~Kang}
\affiliation{Department of Physics, Seoul National University, Seoul 151-742, Korea}

\author{S.~Ajimura}
\affiliation{Department of Physics, Osaka University, Toyonaka 560-0043, Japan}

\author{K.~Aoki}
\affiliation{High Energy Accelerator Research Organization (KEK), Tsukuba 305-0801, Japan}

\author{A.~Banu}
\affiliation{Gesellschaft f\"ur Schwerionenforschung mbH (GSI), Darmstadt 64291, Germany}

\author{H.~Bhang}
\affiliation{Department of Physics, Seoul National University, Seoul 151-742, Korea}

\author{T.~Fukuda}
\affiliation{Laboratory of Physics, Osaka Electro-Communication University, Neyagawa 572-8530, Japan}

\author{O.~Hashimoto}
\affiliation{Physics Department, Tohoku University, Sendai 980-8578, Japan}

\author{J.~I.~Hwang}
\affiliation{Department of Physics, Ewha Womans University, Seoul 120-750, Korea}

\author{S.~Kameoka}
\affiliation{Physics Department, Tohoku University, Sendai 980-8578, Japan}

\author{E.~H.~Kim}
\affiliation{Department of Physics, Seoul National University, Seoul 151-742, Korea}

\author{J.~H.~Kim}
\altaffiliation[Present address: ]{Korea Research Institute of Standards and Science (KRISS), Daejeon, 305-600, Korea.}
\affiliation{Department of Physics, Seoul National University, Seoul 151-742, Korea}

\author{M.~J.~Kim}
\affiliation{Department of Physics, Seoul National University, Seoul 151-742, Korea}

\author{T.~Maruta}
\affiliation{Department of Physics, University of Tokyo, Hongo 113-0033, Japan}

\author{Y.~Miura}
\affiliation{Physics Department, Tohoku University, Sendai 980-8578, Japan}

\author{Y.~Miyake}
\affiliation{Department of Physics, Osaka University, Toyonaka 560-0043, Japan}

\author{T.~Nagae}
\affiliation{High Energy Accelerator Research Organization (KEK), Tsukuba 305-0801, Japan}

\author{M.~Nakamura}
\affiliation{Department of Physics, University of Tokyo, Hongo 113-0033, Japan}

\author{S.~N.~Nakamura}
\affiliation{Physics Department, Tohoku University, Sendai 980-8578, Japan}

\author{H.~Noumi}
\affiliation{High Energy Accelerator Research Organization (KEK), Tsukuba 305-0801, Japan}

\author{S.~Okada}
\altaffiliation[Present address: ]{RIKEN Wako Institute, RIKEN, Wako 351-0198, Japan.}
\affiliation{Department of Physics, Tokyo Institute of Technology, Ookayama 152-8551, Japan}

\author{Y.~Okayasu}
\affiliation{Physics Department, Tohoku University, Sendai 980-8578, Japan}

\author{H.~Outa}
\altaffiliation[Present address: ]{RIKEN Wako Institute, RIKEN, Wako 351-0198, Japan.}
\affiliation{High Energy Accelerator Research Organization (KEK), Tsukuba 305-0801, Japan}

\author{H.~Park}
\affiliation{Korea Research Institute of Standards and Science (KRISS), Daejeon 305-600, Korea}

\author{P.~K.~Saha}
\altaffiliation[Present address: ]{Japan Atomic Energy Research Institute, Tokai 319-1195, Japan.}
\affiliation{High Energy Accelerator Research Organization (KEK), Tsukuba 305-0801, Japan}

\author{Y.~Sato}
\affiliation{High Energy Accelerator Research Organization (KEK), Tsukuba 305-0801, Japan}

\author{M.~Sekimoto}
\affiliation{High Energy Accelerator Research Organization (KEK), Tsukuba 305-0801, Japan}

\author{T.~Takahashi}
\altaffiliation[Present address: ]{High Energy Accelerator Research Organization (KEK), Tsukuba 305-0801, Japan.}
\affiliation{Physics Department, Tohoku University, Sendai 980-8578, Japan}

\author{H.~Tamura}
\affiliation{Physics Department, Tohoku University, Sendai 980-8578, Japan}

\author{K.~Tanida}
\affiliation{RIKEN Wako Institute, RIKEN, Wako 351-0198, Japan}

\author{A.~Toyoda}
\affiliation{High Energy Accelerator Research Organization (KEK), Tsukuba 305-0801, Japan}

\author{K.~Tsukada}
\affiliation{Physics Department, Tohoku University, Sendai 980-8578, Japan}

\author{T.~Watanabe}
\affiliation{Physics Department, Tohoku University, Sendai 980-8578, Japan}

\author{H.~J.~Yim}
\affiliation{Department of Physics, Seoul National University, Seoul 151-742, Korea}

\date{\today}
\begin{abstract}
We performed a coincidence measurement of two nucleons emitted
from the nonmesonic weak decay (NMWD) of $^{5}_{\Lambda}$He formed 
via the $^{6}$Li($\pi^+$,$K^+$) reaction.
The energies of the two nucleons and 
the pair number distributions in the opening angle between them were measured. 
In both $np$ and $nn$ pairs, 
we observed a clean back-to-back correlation coming from the 
two-body weak reactions of 
$\Lambda p \to n p$ and $\Lambda n \to n n$, respectively.
The ratio of the nucleon pair numbers was
$N_{nn}/N_{np}$=0.45$\pm$0.11(stat)$\pm$0.03(syst) 
in the kinematic region of $\cos\theta_{NN} < -0.8$.
Since each decay mode was exclusively detected,
the measured ratio should be close to the ratio of
$\Gamma(\Lambda p\to np)$/$\Gamma(\Lambda n\to nn)$.
 The ratio is consistent with recent theoretical calculations
based on the heavy meson/direct quark exchange picture.
\end{abstract}


\maketitle
A free $\Lambda$ hyperon decays almost totally into a pion and a nucleon, which is a mesonic weak
decay process (\LNpi; $\Delta q \sim 100 MeV/c$). However, 
a $\Lambda$ hyperon bound in a nucleus will eventually 
decay through either a mesonic or a nonmesonic weak decay process 
(\LN;  $\Delta q \sim 400 MeV/c$). 
NMWD, in which a $\Lambda$ decays via a weak interaction
with neighboring nucleon(s), 
becomes possible only inside the nucleus.
The simplest and long assumed NMWD modes have been 
the one-nucleon (1$N$) induced ones, 
namely $\Lambda p\to np$ and 
$\Lambda n \to nn$, whose partial decay widths are denoted as
\Gp~and \Gn, respectively. 
The study of NMWD is of fundamental importance, since it provides a
unique way of exploring the strangeness-changing,
baryon-baryon weak interaction.
In addition, the two-nucleon (2$N$) induced NMWD ($\Lambda NN \to nNN$; \Gnn)
has been predicted in theoretical calculations, 
although its experimental identification has not yet been achieved. 

There has been a long standing "puzzle" concerning the \Gnp~ratio of 
NMWD of $\Lambda$ hypernuclei. 
Until a few years ago the experimental ratios 
had been reported to be close to or 
larger than unity, while the most naive one-pion-exchange
(OPE) model with $\Delta I$=1/2 rule predicts very small
ratios of 0.05$-$0.20.  
  Other decay mechanisms beyond the OPE model have been considered
to explain the large $\Gamma_n/\Gamma_p$ ratio.
 The most relevant are:
1) The heavy-meson-exchange (HME) mechanism in which exchanges of 
   mesons heavier than pions 
   (especially kaons) are considered,
2) The direct-quark exchange (DQ) mechanism where the short range
   $\Lambda N \rightarrow NN$ decay is described
   using quark degrees of freedom, violating
    the $\Delta I=1/2$ rule, and in a different context
3) the inclusion of a two-nucleon induced decay ($\Lambda NN \rightarrow nNN$)
    mechanism.
Refer to the recent review article~\cite{ref:Alb02} for the 
details of various models and their results in terms of NMWD widths. 
  It was recently found that in all previous theoretical
work there had been an error in the K exchange amplitude
and its correction
significantly increased theoretical values of
the \Gnp~ratio.
After the correction was included, the DQ model 
gave 0.70 for the ratio in \He5lam~\cite{ref:Sas00}.
The HME model calculation 
also predicted an increased value of up to 0.34$-$0.46~\cite{ref:Par01}.
 These calculations still underestimate the reported values 
of $\Gamma_n/\Gamma_p$ ratios for light hypernuclei, 
although the large experimental errors (e.g. 0.93$\pm$0.55 for 
$^5_{\Lambda}$He \cite{ref:Szy91}) prevented any definite conclusion being reached.
Accurate measurements of the $\Gamma_n/\Gamma_p$
ratio are still awaited. 

   Concerning the $\Gamma_n$/$\Gamma_p$ ratio, most of the previous 
 experiments  measured energetic protons only and the ratio
 was estimated without any experimental information on neutrons and
 low energy protons
 \cite{ref:Szy91,ref:Has02,ref:Sat03}.
   Thus the result might be much affected by missing the low energy 
 protons caused by the re-scattering process of protons 
 inside the nucleus (Final-State Interaction; FSI)  
 and the possible existence of the 
 two-nucleon induced decay modes.
 Both of these processes may induce the quenching of 
 energetic proton numbers above the detection threshold,
 thus overestimating the $\Gamma_n/\Gamma_p$ ratio.
   Important progress has been made through 
the accurate measurement of neutron spectra 
from the NMWD of \Clam~(the KEK-PS E369 experiment)~\cite{ref:Kim03}.
 A neutron spectrum from the  weak decay of \Clam~was measured with 
an order of magnitude improvement in the 
statistics and signal-to-background ratio compared to the results of
a previous experiment~\cite{ref:Szy91}. 
 This result opened a door to directly comparing the yields of
neutrons to protons from the NMWD.
 Very recently we reported the simultaneously measured  
spectra of neutrons and  protons emitted from NMWD of 
\He5lam~and \Clam~\cite{ref:Oka04} with much 
higher statistics than those of the previous 
experiments~\cite{ref:Szy91,ref:Kim03}.
The neutron to proton yield ratios for both hypernuclei
obtained with high threshold energy (60 MeV) 
are approximately equal to two,
which suggests a \Gnp~ratio of about 0.5, namely 
$\Lambda p \rightarrow n p$ channel dominance.   
However, the results still contained some uncertainties due to 
residual FSI effects and 
a possible contribution from 2$N$-induced NMWD.

\begin{figure}
  \includegraphics[scale=0.38]{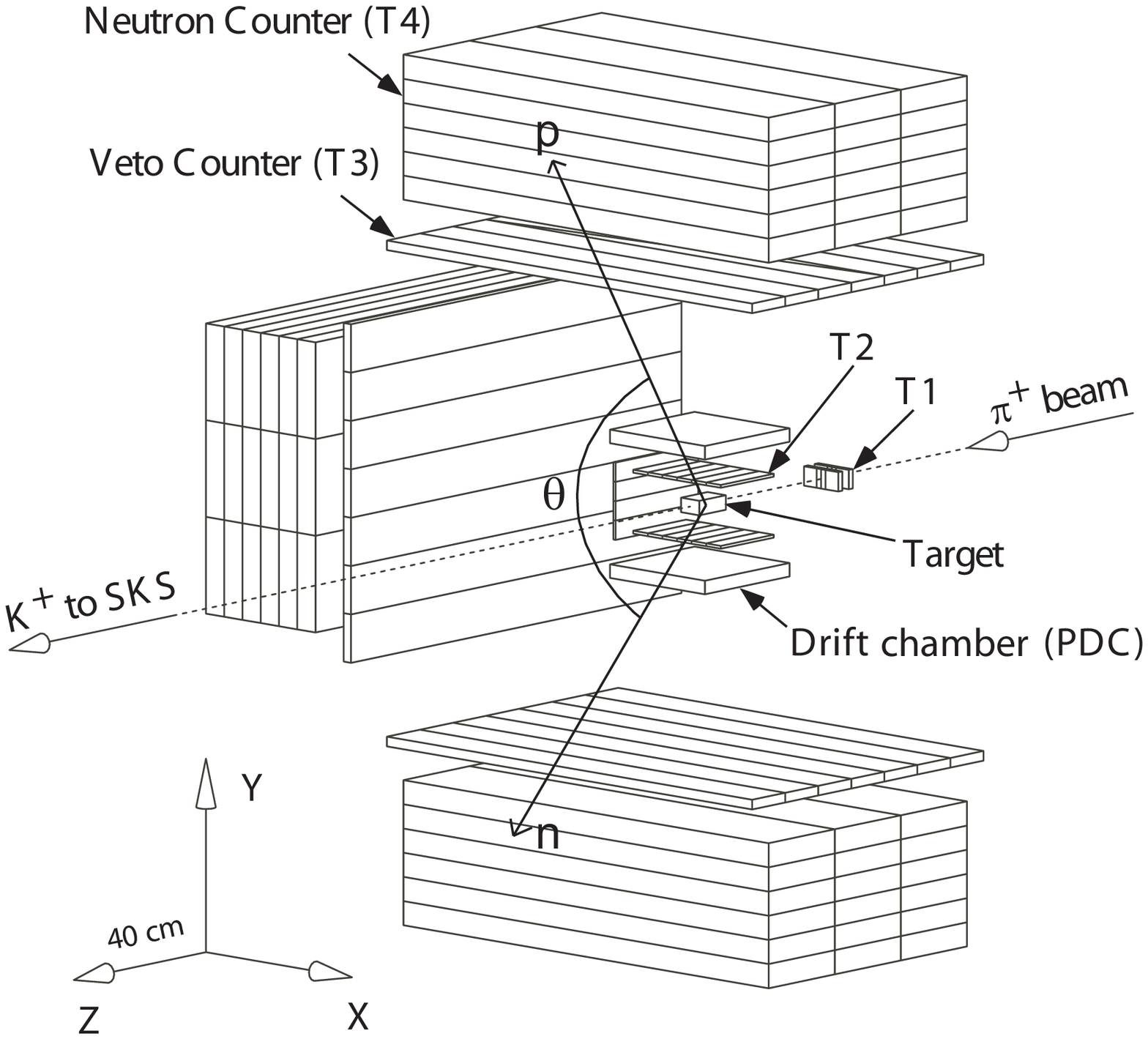}
  \caption{\label{fig:setup} 
    Detector setup for the coincidence measurement of neutrons 
and protons emitted from NMWD in KEK-PS E462.}
\end{figure}

To resolve these experimental difficulties, 
it is important to clearly identify the two-body decay kinematics
of $\Lambda n \to n n$ and $\Lambda p \to n p$.
To identify these decay channels unambiguously, 
we have measured in coincidence two nucleons
emitted back-to-back.
 Here we chose $^5_\Lambda$He hypernuclei because the 
effect of FSI is expected to be small in such light hypernuclei.
The yields of pair coincidences,
$Y_{np}$ and $Y_{nn}$, can be expressed as
\begin{subequations}
\begin{eqnarray}
  \label{eq:NNpair}
    Y_{nn} (cos \theta)&=& Y_{hyp}~ b_{nm}~ r_n~ \e_{nn}(cos \theta)~
       f_n^2, \label{eq:Ynn} \\
    Y_{np} (cos \theta)&=& Y_{hyp}~ b_{nm}~ r_p~ \e_{np}(cos \theta)~ 
       f_n~ f_p, \label{eq:Ynp}
\end{eqnarray}
\end{subequations}
where $Y_{hyp},~b_{nm}$ and $r_{n(p)}$ are the number of \He5lam~produced, 
the branching ratio for NMWD of \He5lam 
and $\Gamma_{n(p)}/\Gamma_{nm}$, respectively.  
The $\e_{np}$ (= $\Omega_{np} \e_n  \e_p$) and $\e_{nn}$(= $\Omega_{nn}  \e_n^2$) are
the overall efficiencies for detecting $np$ and $nn$ pairs
including the detector acceptance
where $\Omega_{NN}$ and $\e_N$ are the detector solid angle 
for the nucleon pair $NN$ and the efficiency for each nucleon $N$.
The $f_{n(p)}$ is the reduction factor due to FSI on the yield of 
neutrons or protons in the energy range above 30 MeV.
Since the 1$N$-induced NMWD is a two-body interaction
and emits nucleons with much higher momentum than
the nuclear Fermi momentum,
one can expect that the two emitted nucleons are strongly 
back-to-back correlated in their opening angle, 
$\theta$, shown in Fig.~\ref{fig:setup}, 
and have their energy-sum $E_{sum}$~close to 
the Q-value (153 MeV) of the NMWD process,
if the $1N$-induced NMWD process occurs without FSI.
We were able to reject 
most of the events due to FSI and 2$N$ processes 
by requiring strict conditions on the  
angular and energy-sum correlation. 
$Y_{nn}$ and $Y_{np}$ are proportional 
to $r_n$ and $r_p$ as shown in Eqs.~\ref{eq:Ynn} and~\ref{eq:Ynp}.

The $r_n/r_p$ ratio, which is identical to the \Gnp~ ratio,
is obtained from the ratio $Y_{nn}/Y_{np}$ as 
\begin{eqnarray}
  \label{eq:gngp}
  \frac{r_n}{r_p} \left( = \frac{N_{nn}}{N_{np}} \right)
  &=&  
  \frac{\{Y_{nn}~/~Y_{hyp}~ b_{nm}~ \epsilon_{nn}\}}
       {\{Y_{np}~/~Y_{hyp}~ b_{nm}~ \epsilon_{np}\}}
       \cdot \frac{f_n~f_p}{f_n^2} \nonumber\\
       &=&  \frac{Y_{nn}~ \Omega_{np}~ \epsilon_p}
             {Y_{np}~ \Omega_{nn}~ \epsilon_n}
\end{eqnarray}
since $Y_{hyp}b_{nm}$ is the total number of NMWD,
$N_{nn}$ and $N_{np}$ are the normalized pair yields 
per NMWD for full solid angle and unit efficiency. 
    We can assume that protons and neutrons suffer the same
 FSI effect in the nucleus, i.e. $f_n = f_p$ 
 because of the charge symmetry of the $NN$ interaction.
By taking the ratio of the $nn$-to-$np$ pair yields, 
many systematic error  sources, like 
$Y_{hyp}$, $b_{nm}$, $\epsilon_n$ and $f_n$, cancel
and we do not need to know the branching ratio of the 2$N$ contribution
to get the \Gnp~ratio from the pair yield ratio.  
In previous experiments, $b_{nm}$ and $f_n$ were sources
of large errors in the \Gnp~ratio.     
 Now we directly obtain the
$\Gamma_n / \Gamma_p$ ratio from experimental quantities.

The experiment (E462) was carried out at the K6 beam line of KEK-PS using a  
1.05 GeV/c $\pi^+$ beam to induce the \Lisix\pik~reaction.
Momenta of both the beam pion and the produced kaon
were measured with $\sigma_p/p\sim 10^{-3}$ resolution 
using the K6 beam spectrometer and
the Superconducting Kaon Spectrometer (SKS). 
They were then used to reconstruct the excitation 
energy spectrum of the hypernucleus produced~\cite{ref:Fuk95}.

The setup detecting the particles from the weak decay of \HefiveL~is 
shown in Fig.~\ref{fig:setup}. It has three sets of decay-particle counters,
two located on the top and bottom of the target 
for a back-to-back coincidence measurement
and one at the side to cover the solid angle for non back-to-back events.
Each of the top and bottom coincidence counter sets consists of fast 
timing counters (T2), a drift chamber (PDC), veto or stop timing counters (T3) 
and neutron counter arrays (T4). 
The side counter is similar but without the PDC.

Neutral particles, $\gamma$'s and neutrons, were measured in 
T4 together with T3 as veto counters.
The particle identification was done in the TOF spectra, 
from T1 to T4.
The charged decay particles were identified 
by the three measured variables, 
$\Delta E$, TOF and $E$ which 
denote the energy loss per unit length measured by the T2 counter, 
the time-of-flight between T2 and T3 and 
the ADC sum of sequentially fired counters, respectively.
Protons and pions were clearly identified, and the pion contamination
in the proton gate was less than 1\%.
The detection thresholds were $\sim$10 MeV for neutrons 
and $\sim$30 MeV for protons.
The neutron detection efficiency of the neutron counter 
system, $\epsilon_n$, was calculated 
by a Monte Carlo simulation with a modified version of the {\small DEMONS}
\cite{ref:Dem92} code, which is applicable to a multi-element neutron detector
and has been tested to reproduce various experimental data~\cite{ref:Kim03}.
For charged particle analysis, only the central segments of T2 and T3 were used 
for more accurate determination of the acceptance and particle identification.
Further details of the decay-particle counters are described 
in Ref.~\cite{ref:Oka04}.
\begin{figure}
  \includegraphics[scale=0.33]{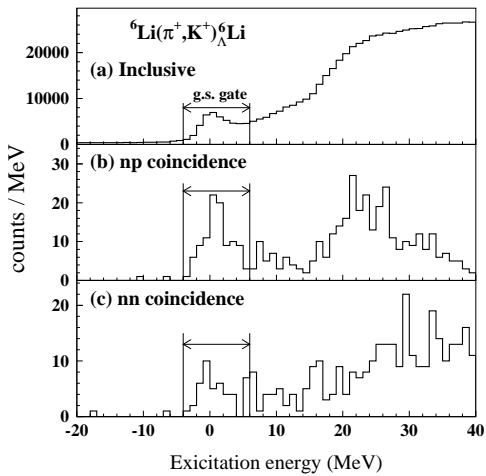}
  \caption{\label{fig:inclusive} 
    The hypernuclear excitation energy spectra for $^{6}_{\Lambda}$Li for
    (a) inclusive, (b) $np$-pair coincidence and (c) $nn$-pair coincidence measurements.}
\end{figure}

The inclusive $^{6}_\Lambda {\rm Li}$ excitation energy spectrum 
derived from the pion and kaon momenta is shown 
in Fig.~\ref{fig:inclusive}(a). 
Figs.~\ref{fig:inclusive}(b) and~\ref{fig:inclusive}(c)
are the spectra in coincidence with nucleon pairs, 
$np$ and $nn$, respectively.
Since the ground state of \Li6lam~is 
above the $^{5}_\Lambda {\rm He} + p$ threshold,
it promptly decays to \HefiveL~by emitting protons of several MeV.
The vertical lines show the
applied gate for the measurement of the decay of $^{5}_\Lambda {\rm He}$.
The enhanced yield in the quasi-free $\Lambda$ region 
of the $nn$ coincidence excitation spectrum over that of 
the $np$ pairs is considered to be due to two-neutron emission 
via the absorption of \pim~from mesonic decay of quasi-free $\Lambda$s.  

\begin{figure}
  \includegraphics[scale=0.56]{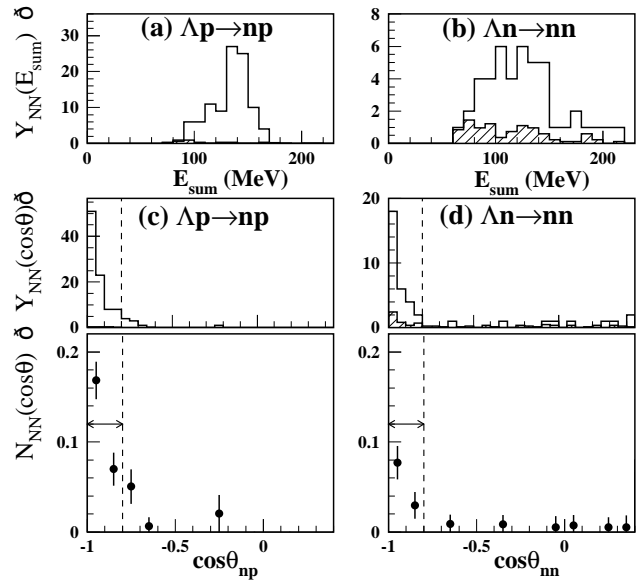}
  \caption{\label{fig:pair} 
    (a) and (b) : The yields of the $np$ and $nn$ coincidence events,
    $Y_{np}$ and $Y_{nn}$,
    as a function of the energy-sum $E_{sum}$ for the pair nucleons.
    The yields are not normalized for efficiency yet. 
    (c) and (d) : The upper panel 
    shows the yields of the $NN$ coincidence
    events, $Y_{NN}$, plotted as a function of the
    the opening angle between two nucleons ($\theta_{NN}$). 
    The lower panel depicts the 
    normalized yields of the $np$ and $nn$ coincidence
    per NMWD.}
\end{figure}

In Figs.~\ref{fig:pair}(a) and \ref{fig:pair}(b), 
the distributions of the $np$ and $nn$ pair yields 
are seen in the energy sum of the pair nucleons.
In the figure only the pair events in which each of the nucleons has
an energy above 30 MeV are counted. 
The hatched histograms 
show the contamination due to the absorption of 
$\pi^-$ emitted from the mesonic decay.   
The amount of contamination
was estimated by referring to the $\pi^-$ absorption 
from mesonic decay of quasi-free $\Lambda$ formation events.
The energy-sum resolution $\sigma_{E_{sum}}$ for $np$ and $nn$~pairs
was estimated 
to be 12 MeV and 16 MeV, respectively, for the typical cases. 
The energy resolution of neutron deteriorates rapidly 
in the high energy region, while that of proton is stable with respect to
its energy.
In the energy-sum spectrum of $np$ pairs, one can see the peak located  
around expected value. 
The sharp peak indicates that the effect of FSI is not severe and 
two-body 1$N$ NMWD is the major process.
Note that the back-to-back kinematic condition has not yet been applied 
to these energy-sum spectra.

The upper panels of Figs.~\ref{fig:pair}(c) and~\ref{fig:pair}(d) 
show the yields of the $np$ and $nn$ 
coincidence events, $Y_{np}$ and $Y_{nn}$, as a function of the opening angle 
between the two nucleons ($\theta_{np}$ and $\theta_{nn}$).
They are not yet normalized for acceptance and efficiency.
The angular resolutions were estimated 
to be 
$\sigma_{\cos(\theta_{np})} = 0.018$ and 
$\sigma_{\cos(\theta_{nn})} = 0.026$
at $\cos\theta = -0.9$.
A total of 90 and 30 events were observed in the back-to-back 
angular region of $\cos\theta < -0.8$ indicated as the vertical 
dashed lines for the $np$ and $nn$ pair yields, respectively.

The lower two panels in Figs.~\ref{fig:pair}(c) and~\ref{fig:pair}(d) show 
the angular correlation of nucleon pairs, where the
vertical scale is normalized for full solid angle and unit efficiency 
($N_{np}$ and $N_{nn}$ in Eq.~\ref{eq:gngp}).
They are normalized with the simulated event-by-event efficiency,
$\e_{NN}(cos \theta)$. 
The $b_{nm}$ value, 0.429$\pm$0.012$\pm$0.005, 
derived from the $\pi^-$ and $\pi^0$ branching ratios measured
in the present experiment~\cite{ref:Oka03,ref:Kam03} and 
drastically improved from the previous 34\% error~\cite{ref:Szy91}.
 This accurate $b_{nm}$~value made it possible to normalize the pair yields 
per NMWD without introducing significant systematic error
so that the measured $N_{NN}$ angular correlation of $nn$ and $np$ pairs
can be directly compared to those of the FSI 
model calculation~\cite{ref:Gar03}.

 Back-to-back peaking at $\cos\theta < -0.8$, 
which is the signature of a two-body final state, is clearly observed 
in the angular correlation for both $np$ and $nn$ pairs.
This is the first clean observation of 
\Lnp~and \Lnn~1$N$-induced NMWD processes.

Next we discuss the
$N_{nn}/N_{np}$ ratio in the back-to-back kinematic region 
($\cos\theta < -0.8$), because the acceptance beyond there decrease rapidly. 
In this kinematic region, the ratio $\Gamma_n /\Gamma_p$ simply becomes 
the $N_{nn}/N_{np}$ ratio as shown in Eq.~\ref{eq:gngp} assuming 
the same FSI on neutrons and protons.
Possible differences in the FSI on neutrons and protons would not be 
significant due to the smallness of the FSI itself in the residual 
nucleus, $^4 \rm{He}$ or $^4 \rm{H}$~\cite{ref:Gar03}. 
The similar spectral shape of the neutron and 
proton single spectra observed in the same experiment supports 
this assumption~\cite{ref:Oka04}. 
 Comparing the yield of $N_{nn}$ and $N_{np}$ in the $cos \theta < -0.8$ region,
we obtained
\begin{eqnarray}
  \label{eq:rnrp}
   \frac{\Gamma_n}{\Gamma_p} 
  \simeq \frac{N_{nn}}{N_{np}}
  = 0.45\pm0.11 (stat) \pm0.03 (syst). 
\end{eqnarray}
The systematic error of about 7\% is mainly from the
ambiguity of neutron detection efficiency, 6\%.
 The \Gnp~ratio is determined 
for the first time from exclusive measurement of each NMWD channel 
by detecting both emitted nucleons in coincidence in order to 
remove the ambiguities of the FSI and 2$N$ NMWD contributions inherent 
in all the previous measurements. 
 This ratio agrees well 
with the recent theoretical ratios of the direct quark interaction
and the heavy meson exchange models~\cite{ref:Sas00,ref:Par01}. 
 Although the $\Gamma_n /\Gamma_p$ ratio is found to be smaller 
than previous values, it is still well
above the OPE prediction.  It is now clear that one needs
additional shorter-range mechanisms, such as DQ and/or HME, 
to explain this strangeness-changing baryon-baryon weak
interaction.

In summary, 
we have measured for the first time the energy-sum and the 
angular correlation of two nucleon pairs, 
$nn$ and $np$, emitted in the NMWD of \He5lam.
We have clearly observed
a distinct $nn$ and $np$ back-to-back correlation
coming from a one-nucleon induced decay mode.
 We have determined the \Gnp~of NMWD of \He5lam
accurately and unambiguously 
from the pair number ratio, $N_{nn}/N_{np}$,
in the back-to-back kinematics region, 
which is almost free from the effects of FSI and 2$N$-induced 
 NMWD contributions.

We are grateful to K.Nakamura and KEK-PS staff for their support 
of our experiment and stable operation of the KEK-PS. 
Authors, BHK and HB, acknowledge 
support from grants KOSEF(R01-2005-000-10050-0) and KRF(2003-070-C00015).

\end{document}